\documentclass[twocolumn,description,aps]{revtex4-1}
\usepackage{graphicx,epstopdf,amsmath,amssymb,multirow,CJK,color}
\usepackage[german,english]{babel}
\usepackage[colorlinks, linkcolor=blue,anchorcolor=blue,citecolor=blue,urlcolor=blue]{hyperref}
\usepackage{booktabs}

\begin{document}

\title{Intrinsic ferromagnetic axion states and a single pair of Weyl fermions in the stable-state Mn\emph{X}$_{2}$\emph{B}$_{2}$\emph{T}$_{6}$-family materials}

\author{Yan Gao$^{1}$}\email{yangao9419@ysu.edu.cn}
\author{Weikang Wu$^{2,3}$}
\author{Ben-Chao Gong$^{4}$}
\author{Huan-Cheng Yang$^{4}$}
\author{Xiang-Feng Zhou$^{1}$}
\author{Yong Liu$^{1}$}
\author{Shengyuan A. Yang$^{3}$}
\author{Kai Liu$^{4}$}
\author{Zhong-Yi Lu$^{4}$}\email{zlu@ruc.edu.cn}

\affiliation{$^{1}$State Key Laboratory of Metastable Materials Science and Technology and Key Laboratory for Microstructural Material Physics of Hebei Province, School of Science, Yanshan University, Qinhuangdao 066004, China}
\affiliation{$^{2}$Key Laboratory for Liquid-Solid Structural Evolution and Processing of Materials (Ministry of Education), Shandong University, Jinan, Shandong 250061, China}
\affiliation{$^{3}$Research Laboratory for Quantum Materials, Singapore University of Technology and Design, Singapore 487372, Singapore}
\affiliation{$^{4}$Department of Physics and Beijing Key Laboratory of Opto-electronic Functional Materials $\&$ Micro-nano Devices, Renmin University of China, Beijing 100872, China}

\date{\today}

\begin{abstract}
The intrinsic ferromagnetic (FM) axion insulators and Weyl semimetals (WSMs) with only single pair of Weyl points have drawn intensive attention but so far remain rare and elusive in real materials. Here, we propose a new class of Mn\emph{X}$_{2}$\emph{B}$_{2}$\emph{T}$_{6}$-B (\emph{X}=Ge, Sn, or Pb; \emph{B}=Sb or Bi; \emph{T}=Se or Te) family that is the stable structural form of this system. We find that the Mn\emph{X}$_{2}$\emph{B}$_{2}$\emph{T}$_{6}$-B family has not only the intrinsic FM axion insulators MnGe$_{2}$Bi$_{2}$Te$_{6}$-B, MnSn$_{2}$Bi$_{2}$Te$_{6}$-B, and MnPb$_{2}$Bi$_{2}$Te$_{6}$-B, but also the intrinsic WSM MnSn$_{2}$Sb$_{2}$Te$_{6}$-B with only a single pair of Weyl points. Thus, the Mn\emph{X}$_{2}$\emph{B}$_{2}$\emph{T}$_{6}$-B family can provide an ideal platform to explore the exotic topological magnetoelectric effect and the intrinsic properties related to Weyl points.
\end{abstract}

\date{\today} \maketitle


$Introduction$: The discovery of the intrinsic antiferromagnetic (AFM) topological insulator (TI) MnBi$_{2}$Te$_{4}$ has attracted intensive attention in recent years~\cite{1Otrokov,2Deng,3Gong,4Li,5Zhang,6Otrokov,7Liu,8Hao,9Zeugner,10Li,11Chen,12LiYan,13Zhang}. On one hand, it has tunability in layer thickness and external magnetic field as well as various novel magnetic topological phases such as the quantum anomalous Hall insulators~\cite{2Deng,4Li}, the axion insulators (AXIs)~\cite{5Zhang,6Otrokov,7Liu}, and the magnetic Weyl semimetals (WSMs)~\cite{4Li,5Zhang}. On the other hand, its unique way of introducing magnetism can provide new insights and vitality in the search for materials with coexisting long-range magnetic order and nontrivial band topology. The MnBi$_{2}$Te$_{4}$ is formed by inserting an electrically neutral [MnTe] atomic layer into its parent TI Bi$_{2}$Te$_{3}$, which avoids the introduction of additional charges of magnetic impurity atoms and only introduces the magnetism of Mn$^{2+}$ ions. But, the intrinsic A-type AFM MnBi$_{2}$Te$_{4}$ can only realize the AFM axion insulator state and the Weyl semimetal state with only a pair of Weyl points (WPs) under external magnetic field~\cite{5Zhang,14Li}. It is worth noting that the ferromagnetic (FM) axion insulators have recently been proposed as an ideal platform to achieve desirable topological magnetoelectric responses~\cite{15Wan}, while the ideal Weyl semimetals~\cite{16Bernevig} with the minimum number of WPs can serve as the simplest template to study the intrinsic physical properties of the WPs and the application of related devices based on the WPs. However, the practical materials with the intrinsic FM axion insulators~\cite{17Hu,18Gao} and intrinsic WSM states with only a pair of Weyl fermions~\cite{19Nie} have so far been very rare.

\begin{figure}[th]
	\centering
	\includegraphics[width=0.44\textwidth]{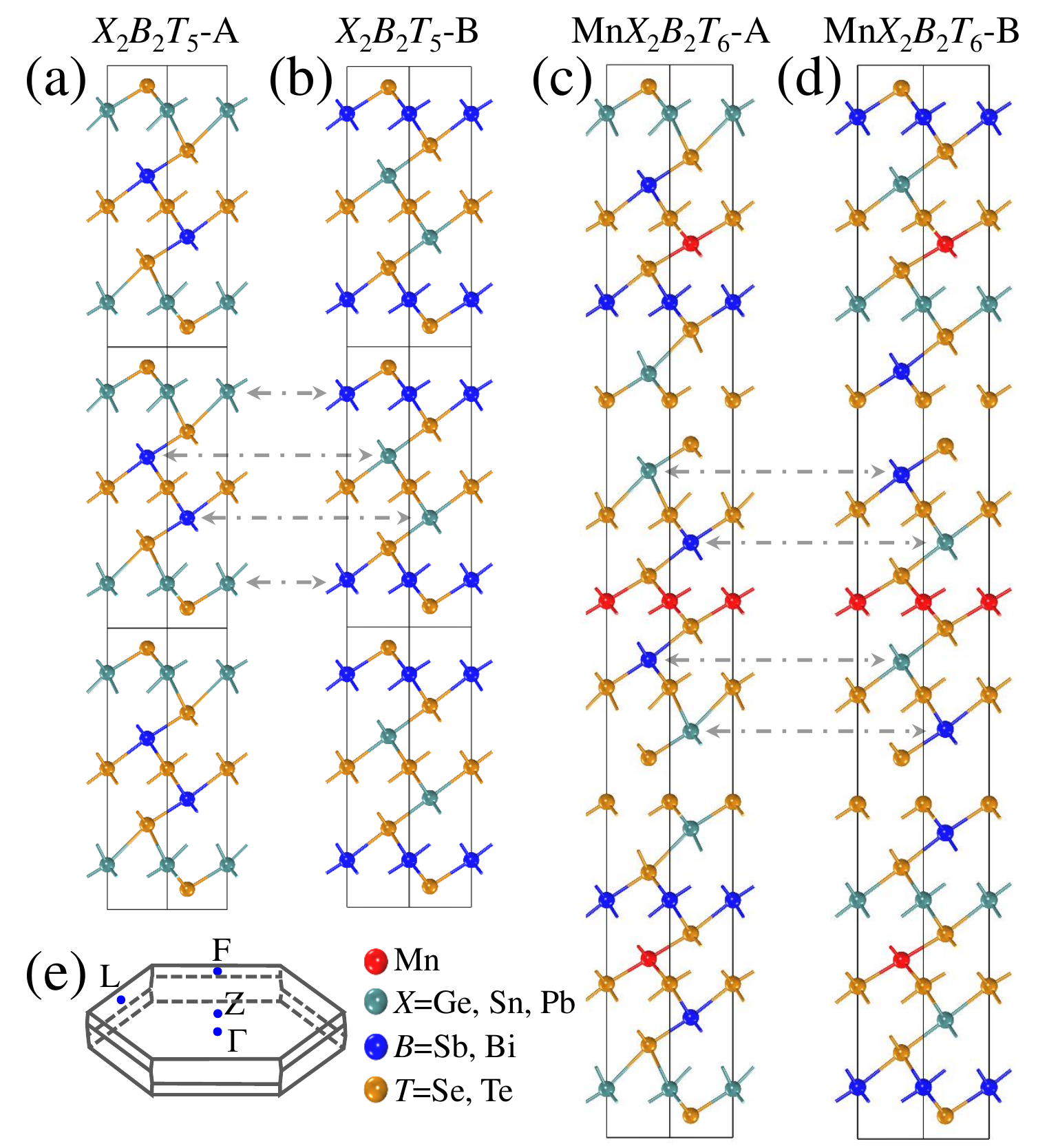}
	\caption{(Color online) Optimized crystal structures of the Mn\emph{X}$_{2}$\emph{B}$_{2}$\emph{T}$_{6}$ (\emph{X}=Ge, Sn, or Pb; \emph{B}=Sb or Bi; \emph{T}=Se or Te) and its parent \emph{X}$_{2}$\emph{B}$_{2}$\emph{T}$_{5}$ materials. The $1 \times 1 \times 3$ supercells of (a) \emph{X}$_{2}$\emph{B}$_{2}$\emph{T}$_{5}$-A and (b) \emph{X}$_{2}$\emph{B}$_{2}$\emph{T}$_{5}$-B in stacking sequences A and B, respectively. Conventional cells of (c) Mn\emph{X}$_{2}$\emph{B}$_{2}$\emph{T}$_{6}$-A and (d) Mn\emph{X}$_{2}$\emph{B}$_{2}$\emph{T}$_{6}$-B in stacking sequences A and B, respectively. (e) Brillouin zone (BZ) of the primitive cell of Mn\emph{X}$_{2}$\emph{B}$_{2}$\emph{T}$_{6}$-B.}
	\label{fig_structure}
\end{figure}

Inspired by MnBi$_{2}$Te$_{4}$, in our previous work we designed a class of magnetic axion insulators Mn\emph{X}$_{2}$\emph{B}$_{2}$\emph{T}$_{6}$-A (\emph{X}=Ge, Sn, or Pb; \emph{B}=Sb or Bi; \emph{T}=Se or Te) family~\cite{18Gao}, which is composed of the parent TI \emph{X}$_{2}$\emph{B}$_{2}$\emph{T}$_{5}$-A family and [MnTe] atomic intercalation. However, we found that the parent material \emph{X}$_{2}$\emph{B}$_{2}$\emph{T}$_{5}$ family possesses two distinct stacking phases. Taking Pb$_{2}$Bi$_{2}$Te$_{5}$ as an example, both experiments~\cite{20Petrov,21Chatterjee} and theory~\cite{22Ma,23Silkin} confirm that it has two different stacking sequences A (-Te-Pb-Te-Bi-Te-Bi-Te-Pb-Te-) and B (-Te-Bi-Te-Pb-Te-Pb-Te-Bi-Te-), labeled as Pb$_{2}$Bi$_{2}$Te$_{5}$-A and Pb$_{2}$Bi$_{2}$Te$_{5}$-B, respectively, and the \emph{X}$_{2}$\emph{B}$_{2}$\emph{T}$_{5}$ and Mn\emph{X}$_{2}$\emph{B}$_{2}$\emph{T}$_{6}$ families also adopt similar labels to distinguish these two different stacking sequences. We can see that the difference between \emph{X}$_{2}$\emph{B}$_{2}$\emph{T}$_{5}$-A and \emph{X}$_{2}$\emph{B}$_{2}$\emph{T}$_{5}$-B is that the \emph{X} (Ge, Sn, or Pb) and \emph{B} (Sb or Bi) atoms have exchanged positions with each other [see Figs.~\ref{fig_structure}(a) and~\ref{fig_structure}(b)], so the [\emph{XT}] atomic layer of \emph{X}$_{2}$\emph{B}$_{2}$\emph{T}$_{5}$-A is in the outermost layer, while the [\emph{XT}] atomic layer of \emph{X}$_{2}$\emph{B}$_{2}$\emph{T}$_{5}$-B in the innermost layer. Previous studies~\cite{22Ma} have shown that Pb$_{2}$Bi$_{2}$Te$_{5}$-B is the stable structural form of Pb$_{2}$Bi$_{2}$Te$_{5}$-A in the range of temperatures 1000~K, and the two stacking phases show significant differences in electronic properties. Then one may naturally ask whether the Mn\emph{X}$_{2}$\emph{B}$_{2}$\emph{T}$_{6}$-B family obtained by inserting the [MnTe] atomic layer into the stable parent TI \emph{X}$_{2}$\emph{B}$_{2}$\emph{T}$_{5}$-B material is the stable structure with lower energy than the Mn\emph{X}$_{2}$\emph{B}$_{2}$\emph{T}$_{6}$-A family? Could there be a new topological phase in Mn\emph{X}$_{2}$\emph{B}$_{2}$\emph{T}$_{6}$-B due to the structural difference between the Mn\emph{X}$_{2}$\emph{B}$_{2}$\emph{T}$_{6}$-B and Mn\emph{X}$_{2}$\emph{B}$_{2}$\emph{T}$_{6}$-A families?

In this work, we systematically investigate the stability, magnetic, electronic and topological properties of the Mn\emph{X}$_{2}$\emph{B}$_{2}$\emph{T}$_{6}$-B family by first-principles electronic structure calculations. Our calculations indicate that the Mn\emph{X}$_{2}$\emph{B}$_{2}$\emph{T}$_{6}$-B family has indeed a stable structure with lower energy than the Mn\emph{X}$_{2}$\emph{B}$_{2}$\emph{T}$_{6}$-A family. Surprisingly, we find not only a series of intrinsic ferromagnetic (FM) axion insulators MnGe$_{2}$Bi$_{2}$Te$_{6}$-B, MnSn$_{2}$Bi$_{2}$Te$_{6}$-B, and MnPb$_{2}$Bi$_{2}$Te$_{6}$-B, but also an intrinsic Weyl semimetal (WSM) MnSn$_{2}$Sb$_{2}$Te$_{6}$-B with only a pair of Weyl fermions, which has not been reported in the Mn\emph{X}$_{2}$\emph{B}$_{2}$\emph{T}$_{6}$-A family. We further investigate the effect of lattice strain on the topological properties of the MnSn$_{2}$Sb$_{2}$Te$_{6}$-B bulk materials. Compared with the MnBi$_{2}$Te$_{4}$ and the Mn\emph{X}$_{2}$\emph{B}$_{2}$\emph{T}$_{6}$-A family, the Mn\emph{X}$_{2}$\emph{B}$_{2}$\emph{T}$_{6}$-B family can provide a more ideal platform for future experiments to explore the unique topological quantum physics of the FM axion insulators and WSMs.

The electronic structures of the Mn\emph{X}$_{2}$\emph{B}$_{2}$\emph{T}$_{6}$-B (\emph{X}=Ge, Sn, or Pb; \emph{B}=Sb or Bi; \emph{T}=Se or Te) family were investigated with the projector augmented wave method~\cite{24PBE} as implemented in the VASP package~\cite{25Kresse} in the framework of density functional theory (DFT). The generalized gradient approximation (GGA) of the Perdew-Burke-Ernzerhof (PBE) type~\cite{26Perdew} was adopted for the exchange-correlation functional. The zero damping DFT-D3 method~\cite{27Grimme} was chosen for the interlayer vdW interaction. The PBE+$U$ method was utilized to treat the localized 3\emph{d} orbitals of Mn by selecting $U$= 4.0~eV. The $6 \times 6 \times 6$, $8 \times 8 \times 8$ and $12 \times 12 \times 1$ k-point meshes~\cite{28Monkhorst} were adopted for the primitive cell calculations of the AFM bulks, FM bulks, and the monolayers, respectively. An 18~\text{\AA} vacuum layer was used to avoid the residual interactions between neighboring image layers. The energy and force convergence criteria were set to $10^{-6}$~eV and 0.001~eV/\text{\AA}, respectively. The phonon spectrum calculations were performed by using the DS-PAW software integrated in the Device Studio program~\cite{29Hongzhiwei}, in which the $4 \times 4 \times 1$ supercell was used for the monolayer structures. The topological properties of the Mn\emph{X}$_{2}$\emph{B}$_{2}$\emph{T}$_{6}$-B family were calculated by using the Wannier90~\cite{30Mostofi} and WannierTools~\cite{31Wu} packages. An \emph{ab} \emph{initio} evolutionary algorithm, as implemented in the USPEX code~\cite{32Oganov,33Lyakhov}, was employed to search for the stable bulk compounds of the MnPb$_{2}$Bi$_{2}$Te$_{6}$ system.

\begin{table*}[!th]
\caption{\label{tab:I} The crystal structure information and magnetic properties of seven Mn\emph{X}$_{2}$\emph{B}$_{2}$\emph{T}$_{6}$-B monolayers, including the in-plane lattice constants optimized in the respective ferromagnetic (FM) ground state, the bond angles (Mn-\emph{T}-Mn) connecting two nearest Mn atoms, the energy differences ($\Delta{E}$) of various typical magnetic configurations (the nonmagnetic (NM), FM, single-row AFM (SR-AFM), and double-row AFM (DR-AFM) states) with respect to the FM state in the absence of SOC, the energy differences ($\Delta{FM}_{\rm{B-A}}$) between the FM Mn\emph{X}$_{2}$\emph{B}$_{2}$\emph{T}$_{6}$-B and the FM Mn\emph{X}$_{2}$\emph{B}$_{2}$\emph{T}$_{6}$-A monolayers, the local magnetic moments (\emph{M}$_{\rm{a}}$), and the energy difference (magnetic anisotropy energy (MAE)) between the in-plane and out-of-plane FM states calculated with the SOC.}
\begin{center}
\begin{tabular*}{2.0\columnwidth}{@{\extracolsep{\fill}}cccccccccc}
\hline\hline
\multirow{3}{*}{Monolayers}  & \multirow{2}{*}{Lattice constants} & \multirow{2}{*}{Angles ($^{\circ}$)} & \multicolumn{5}{c}{Without SOC} & \multicolumn{2}{c}{\multirow{2}{*}{With SOC}} \\
                             &                                       &                            & \multicolumn{5}{c}{$\Delta{E}$ (meV/Mn)}  & \multicolumn{2}{c}{}                          \\
\cline{4-8}
                             & {$a=b$ (\text{\AA})}                                   & Mn-\emph{T}-Mn                    & NM      & FM  & SR-AFM & DR-AFM & $\Delta{FM}_{\rm{B-A}}$    & {\emph{M}$_{\rm{a}}$ ($\mu_{\rm{B}}$)}     &  MAE   (meV)    \\
\hline
MnGe$_{2}$Sb$_{2}$Se$_{6}$-B &  4.03                                   & 92.7                      & 4357.2  & 0.0 & 3.1    & 0.8  & -186.1  & 4.56         & 0.04           \\
MnGe$_{2}$Sb$_{2}$Te$_{6}$-B &  4.28                                   & 92.0                       & 4312.8  & 0.0 & 4.7    & 2.0  & -73.4  & 4.54         & 0.03           \\
MnGe$_{2}$Bi$_{2}$Se$_{6}$-B &  4.08                                   & 93.7                       & 4410.2  & 0.0 & 3.2    & 1.0  & -212.0  & 4.57         & 0.02           \\
MnGe$_{2}$Bi$_{2}$Te$_{6}$-B &  4.33                                   & 92.8                       & 4365.1  & 0.0 & 3.6    & 1.7  & -98.3  & 4.55         & 0.04           \\
MnSn$_{2}$Sb$_{2}$Te$_{6}$-B &  4.38                                   & 93.8                       & 4301.6  & 0.0 & 4.8    & 2.2  & -154.6 & 4.56         & 0.11           \\
MnSn$_{2}$Bi$_{2}$Te$_{6}$-B &  4.43                                   & 94.7                       & 4358.8  & 0.0 & 4.2    & 1.8  & -185.2  & 4.56         & 0.01           \\
MnPb$_{2}$Bi$_{2}$Te$_{6}$-B &  4.46                                   & 95.2                       & 4414.9  & 0.0 & 2.4    & 0.4  & -505.2  & 4.57         & 0.08           \\
\hline\hline
\end{tabular*}
\end{center}
\end{table*}


$Results$: To investigate the structural stability of the Mn\emph{X}$_{2}$\emph{B}$_{2}$\emph{T}$_{6}$-B family, we have calculated the phonon spectra of the Mn\emph{X}$_{2}$\emph{B}$_{2}$\emph{T}$_{6}$-B monolayers and bulks. As shown in Figs. S1 and S2 of Supplemental Material (SM), there are no soft phonon modes in the whole Brillouin zone (BZ), indicating that MnGe$_{2}$Sb$_{2}$Se$_{6}$-B, MnGe$_{2}$Sb$_{2}$Te$_{6}$-B, MnGe$_{2}$Bi$_{2}$Se$_{6}$-B, MnGe$_{2}$Bi$_{2}$Te$_{6}$-B, MnSn$_{2}$Sb$_{2}$Te$_{6}$-B, MnSn$_{2}$Bi$_{2}$Te$_{6}$-B, and MnPb$_{2}$Bi$_{2}$Te$_{6}$-B are dynamically stable. Next, we mainly focus on these seven Mn\emph{X}$_{2}$\emph{B}$_{2}$\emph{T}$_{6}$-B bulk compounds.

Considering that the interlayer Mn-Mn distance in the Mn\emph{X}$_{2}$\emph{B}$_{2}$\emph{T}$_{6}$-B bulk is very large ($\sim$20~\text{\AA}) and interrupted by the vdW gap, we first consider the various magnetic structures in their monolayers (see Fig. 1 in Ref.~\cite{18Gao}) and then check the interlayer magnetic coupling to determine the magnetic ground states of the Mn\emph{X}$_{2}$\emph{B}$_{2}$\emph{T}$_{6}$-B compounds. In our previous work on Mn\emph{X}$_{2}$\emph{B}$_{2}$\emph{T}$_{6}$-A, various Hubbard $U$ values (such as $U$=3, 4, and 5~eV) on the Mn 3\emph{d} orbitals were tested and we found that these selected $U$ values had little effect on the structural, magnetic, and electronic properties. Thus, in this work, we adopt $U$=4~eV as in Refs.~\cite{4Li,34Li}. From Table \ref{tab:I}, we can see that the energy of nonmagnetic (NM) state is much higher than those of other magnetic configurations, indicating that these structures all have magnetic interactions. Further, our calculations show that the FM order has the lowest energy among various magnetic configurations, suggesting that the magnetic ground state of Mn\emph{X}$_{2}$\emph{B}$_{2}$\emph{T}$_{6}$-B monolayers is the FM state with an out-of-plane easy magnetization axis. Moreover, the seven Mn\emph{X}$_{2}$\emph{B}$_{2}$\emph{T}$_{6}$-B monolayers in the FM ground state all exhibit good FM semiconductor properties in the presence of spin-orbit coupling (SOC) [see Fig. S4 in the SM], which may have potential applications in spintronic devices~\cite{35MacDonald,36Sarma}.

\begin{figure}[th]
	\centering
	\includegraphics[width=0.40\textwidth]{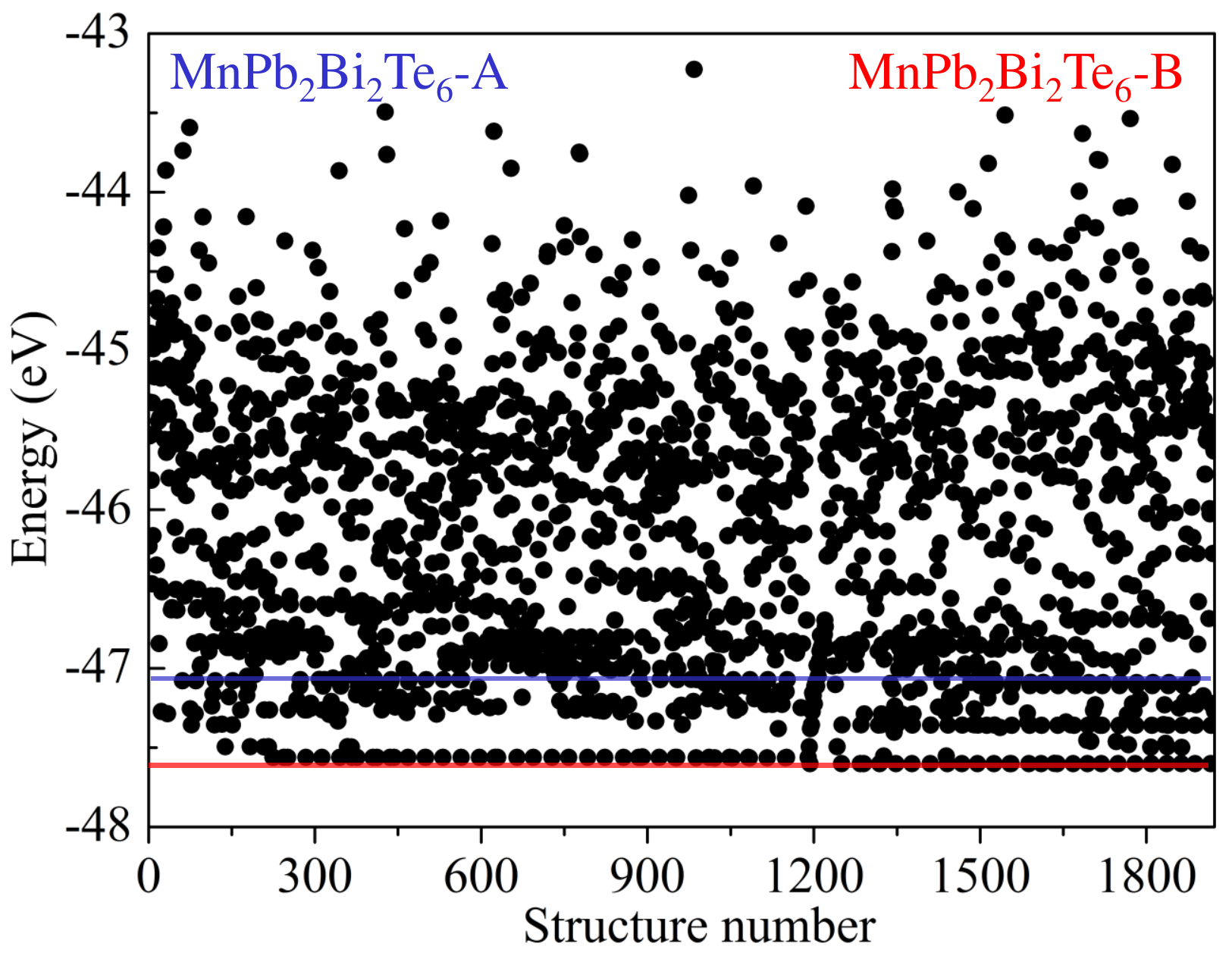}
	\caption{(Color online) Total energy evolution for the MnPb$_{2}$Bi$_{2}$Te$_{6}$ system during the magnetic evolutionary search. The red and blue lines represent the most stable structure MnPb$_{2}$Bi$_{2}$Te$_{6}$-B and metastable MnPb$_{2}$Bi$_{2}$Te$_{6}$-A, respectively.}
	\label{fig_USPEX}
\end{figure}

The crystal structures of Mn\emph{X}$_{2}$\emph{B}$_{2}$\emph{T}$_{6}$-B [Fig.~\ref{fig_structure}(d)] and Mn\emph{X}$_{2}$\emph{B}$_{2}$\emph{T}$_{6}$-A [Fig.~\ref{fig_structure}(c)] bulks are very similar and have the same space group \emph{D}$_{3d}^5$ (No. 166), both of which are formed by ABC stacking of 11 atomic-layers building blocks along the \emph{c}-axis via the vdW interaction. However, their difference in the structure is that the \emph{X} (Ge, Sn, or Pb) and \emph{B} (Sb or Bi) atoms exchange positions with each other. Obviously, this is derived from the difference between their parent \emph{X}$_{2}$\emph{B}$_{2}$\emph{T}$_{5}$-B [Fig.~\ref{fig_structure}(b)] and \emph{X}$_{2}$\emph{B}$_{2}$\emph{T}$_{5}$-A [Fig.~\ref{fig_structure}(a)] structures. For instance, Pb$_{2}$Bi$_{2}$Te$_{5}$-A and Pb$_{2}$Bi$_{2}$Te$_{5}$-B, which have been prepared experimentally~\cite{20Petrov,21Chatterjee}, exchange the positions of Pb and Bi atoms. It is worth noting that the Pb$_{2}$Bi$_{2}$Te$_{5}$-B is a stable structural form of Pb$_{2}$Bi$_{2}$Te$_{5}$-A in the range of temperatures 1000~K~\cite{22Ma}. Therefore, we focus on the stable parent \emph{X}$_{2}$\emph{B}$_{2}$\emph{T}$_{5}$-B structures. A natural question is whether the Mn\emph{X}$_{2}$\emph{B}$_{2}$\emph{T}$_{6}$-B family derived from this stable parent \emph{X}$_{2}$\emph{B}$_{2}$\emph{T}$_{5}$-B can have a stable structural form with lower energy than that of the Mn\emph{X}$_{2}$\emph{B}$_{2}$\emph{T}$_{6}$-A family? To this end, we firstly searched the MnBi$_{2}$Te$_{4}$ system based on the USPEX evolutionary algorithm and quickly obtained the real synthesized MnBi$_{2}$Te$_{4}$ crystals~\cite{37Lee} with the space group \emph{R}$\bar{3}$\emph{m} [see Fig. S3 in the SM], which confirmed the effectiveness of the method. After that, we carried out a systematic search for the MnPb$_{2}$Bi$_{2}$Te$_{6}$ system, and found that the energy of MnPb$_{2}$Bi$_{2}$Te$_{6}$-B phase is indeed lower than that of the MnPb$_{2}$Bi$_{2}$Te$_{6}$-A phase~\cite{18Gao} by nearly 526.9~meV/Mn (see Table \ref{tab:II}). The MnPb$_{2}$Bi$_{2}$Te$_{6}$-B phase is located at the lowest energy (Fig.~\ref{fig_USPEX}), showing that the MnPb$_{2}$Bi$_{2}$Te$_{6}$-B phase is the ground state of the MnPb$_{2}$Bi$_{2}$Te$_{6}$ system.

\begin{table*}[!th]
\caption{\label{tab:II} The structure information, magnetic, and topological properties of seven bulk compounds of the Mn\emph{X}$_{2}$\emph{B}$_{2}$\emph{T}$_{6}$-B family, including the lattice constants optimized in their respective ground states, the bond angles (Mn-\emph{T}-Mn) connecting two nearest Mn atoms, the local magnetic moments (\emph{M}$_{\rm{a}}$), the energy differences ($\Delta{E}_{\rm{B-A}}$) of the Mn\emph{X}$_{2}$\emph{B}$_{2}$\emph{T}$_{6}$-B phase with respect to the Mn\emph{X}$_{2}$\emph{B}$_{2}$\emph{T}$_{6}$-A phase, the energy differences ($\Delta{E}_{\rm{AFM-FM}}$) between the A-type AFM and FM orders, the energy differences (magnetic anisotropy energy (MAE)) between the in-plane and out-of-plane spin orientations in their respective ground states, the number of occupied bands with even and odd parity eigenvalues ($n_{occ}^+$,$n_{occ}^-$) at the eight inversion-invariant momentum points ($\Lambda_{\alpha}$) in the presence of SOC, and the parity-based Z$_{4}$ invariant.}
\begin{center}
\begin{tabular*}{2.07\columnwidth}{@{\extracolsep{\fill}}ccccccccccccc}
\hline\hline
\multirow{2}{*}{Bulks} & \multicolumn{2}{c}{Lattice constants} & {Angles ($^{\circ}$)} & {\emph{M}$_{\rm{a}}$} & {$\Delta{E}_{\rm{B-A}}$} & {$\Delta{E}_{\rm{AFM-FM}}$} & MAE & \multicolumn{4}{c}{$\Lambda_{\alpha}$} & \multirow{2}{*}{Z$_{4}$} \\
\cline{2-3}\cline{9-12}
  & {\emph{a}=\emph{b} (\text{\AA})} & {\emph{c} (\text{\AA})} & {Mn-\emph{T}-Mn} & ($\mu_{\rm{B}}$) & \multicolumn{2}{c}{meV/Mn} & (meV) & $\Gamma$ (0,0,0) & L (0,$\pi$,0) & F ($\pi$,$\pi$,0) & Z ($\pi$,$\pi$,$\pi$) &  \\
\hline
 MnGe$_{2}$Sb$_{2}$Se$_{6}$-B & 4.00 & 57.69 & 93.0 & 4.55 & -149.0 & -0.097 & 0.41 & (86,88) & (87,87) & (86,88) & (87,87) & 0 \\
 MnGe$_{2}$Sb$_{2}$Te$_{6}$-B & 4.27 & 60.91 & 92.4 & 4.53 & -56.6 & 0.043 & 0.17 & (43,44) & (43,44) & (43,44) & (43,44) & 0 \\
 MnGe$_{2}$Bi$_{2}$Se$_{6}$-B & 4.06 & 58.17 & 94.0 & 4.56 & -131.6 & -0.095 & 0.02 & (106,108) & (107,107) & (106,108) & (107,107) & 0 \\
 MnGe$_{2}$Bi$_{2}$Te$_{6}$-B & 4.34 & 61.68 & 93.9 & 4.54 & -42.9 & 0.038 & 0.12 & (55,52) & (53,54) & (53,54) & (53,54) & 2 \\
 MnSn$_{2}$Sb$_{2}$Te$_{6}$-B & 4.36 & 62.05 & 94.4 & 4.54 & -193.4 & 0.14 & 0.10 & (44,43) & (43,44) & (43,44) & (43,44) & 1 \\
 MnSn$_{2}$Bi$_{2}$Te$_{6}$-B & 4.42 & 62.46 & 95.2 & 4.55 & -196.2 & 0.125 & 0.43 & (55,52) & (53,54) & (53,54) & (53,54) & 2 \\
 MnPb$_{2}$Bi$_{2}$Te$_{6}$-B & 4.44 & 63.20 & 95.9 & 4.56 & -526.9 & 0.035 & 0.29 & (55,52) & (53,54) & (53,54) & (53,54) & 2 \\
\hline\hline
\end{tabular*}
\end{center}
\end{table*}

After confirming the intralayer FM ground state of Mn\emph{X}$_{2}$\emph{B}$_{2}$\emph{T}$_{6}$-B monolayer, we only need to consider the interlayer magnetic interaction to determine the magnetic ground state of Mn\emph{X}$_{2}$\emph{B}$_{2}$\emph{T}$_{6}$-B bulks. From Table \ref{tab:II}, we can see that MnGe$_{2}$Sb$_{2}$Se$_{6}$-B and MnGe$_{2}$Bi$_{2}$Se$_{6}$-B bulk exhibit the A-type AFM magnetic ground state similar to MnBi$_{2}$Te$_{4}$~\cite{3Gong,4Li,5Zhang}, while the remaining five Mn\emph{X}$_{2}$\emph{B}$_{2}$\emph{T}$_{6}$-B bulks all have the FM ground state. From the magnetic anisotropy energy (MAE), the seven bulk compounds all host the same out-of-plane easy axis of magnetization as their monolayers. Meanwhile, we note that the energies of the seven Mn\emph{X}$_{2}$\emph{B}$_{2}$\emph{T}$_{6}$-B bulks are 42.9-526.9~meV/Mn lower than those of the Mn\emph{X}$_{2}$\emph{B}$_{2}$\emph{T}$_{6}$-A bulks in their respective ground states, suggesting that Mn\emph{X}$_{2}$\emph{B}$_{2}$\emph{T}$_{6}$-B is the stable structural form of the Mn\emph{X}$_{2}$\emph{B}$_{2}$\emph{T}$_{6}$ system.

\begin{figure}[!th]
	\centering
	\includegraphics[width=0.35\textwidth]{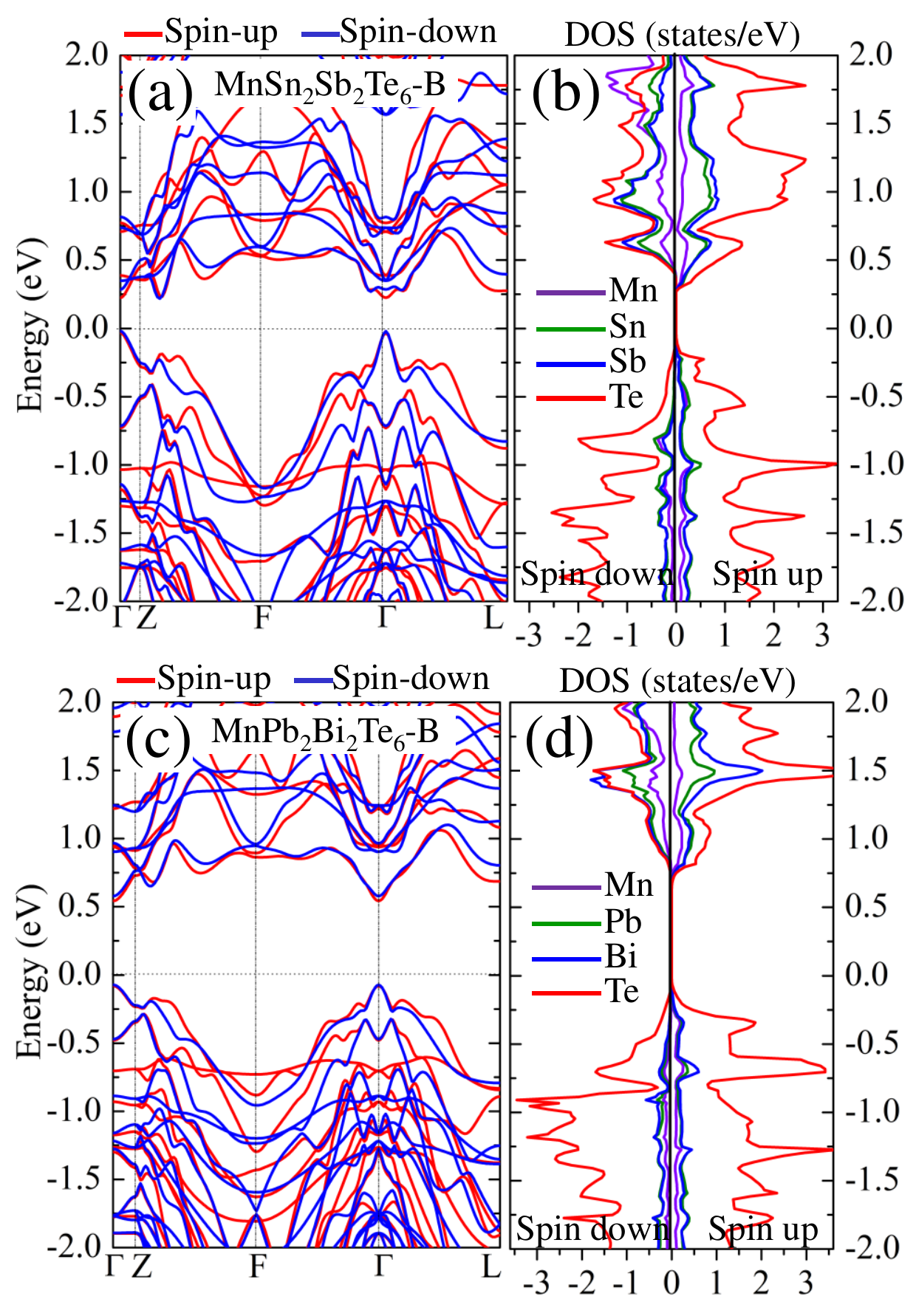}
	\caption{(Color online) Spin-polarized band structures and density of states (DOS) of the bulk compounds MnSn$_{2}$Sb$_{2}$Te$_{6}$-B and MnPb$_{2}$Bi$_{2}$Te$_{6}$-B calculated without the SOC in the ferromagnetic ground state.}
	\label{fig_noSOCband}
\end{figure}

The most interesting property of the Mn\emph{X}$_{2}$\emph{B}$_{2}$\emph{T}$_{6}$-B bulks is its electronic band structure with the abundant topological phases. Figure~\ref{fig_noSOCband} shows the spin-polarized band structures and density of states (DOS) of MnSn$_{2}$Sb$_{2}$Te$_{6}$-B and MnPb$_{2}$Bi$_{2}$Te$_{6}$-B bulks in the FM ground state. We can find that they both are FM semiconductors with narrow band gaps (0.23 and 0.61~eV), and their energy bands near the top of valence bands mainly come from the contributions of the \emph{p} orbitals of Te, Sb/Bi, and Sn/Pb atoms. The spin-polarized band structures of the other five Mn\emph{X}$_{2}$\emph{B}$_{2}$\emph{T}$_{6}$-B bulks in their respective ground states are shown in Fig. S5 of the SM.

\begin{figure}[!th]
	\centering
	\includegraphics[width=0.47\textwidth]{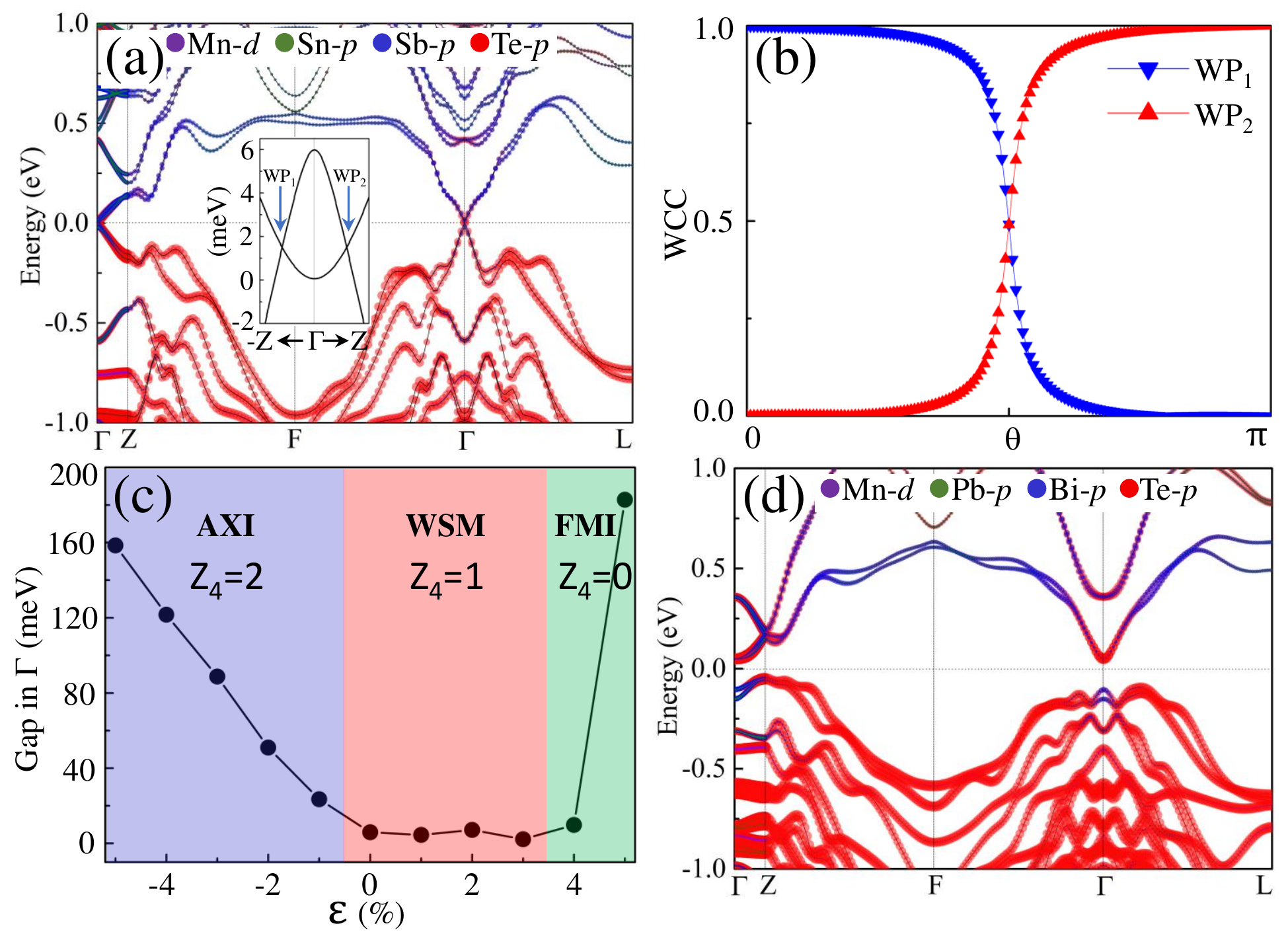}
	\caption{(Color online) Band structures with orbital weights for (a) MnSn$_{2}$Sb$_{2}$Te$_{6}$-B and (d) MnPb$_{2}$Bi$_{2}$Te$_{6}$-B bulk materials calculated with the SOC in the out-of-plane FM ground states. Single pair of Weyl points along the -Z-$\Gamma$-Z path near the Fermi level are shown in inset (a). Two ideal Weyl points (WP$_{1}$ and WP$_{2}$) are separated ~0.06 ${\text{\AA}}^{-1}$ along the \emph{k}$_{z}$ direction, which is comparable to MnBi$_{2}$Te$_{4}$ in the FM state. (b) Motions of the sum of Wannier charge centers (WCCs) on a sphere enclosing each Weyl point in the BZ. (c) The energy gap between the lowest unoccupied state and the highest occupied state at $\Gamma$ as a function of stress along the \emph{c}-axis.}
	\label{fig_SOCband}
\end{figure}

When the SOC effect is taken into account, the band structure of MnSn$_{2}$Sb$_{2}$Te$_{6}$-B bulk in the FM ground state with an out-of-plane easy magnetization axis is shown in Fig.~\ref{fig_SOCband}(a). One can see that it has only a minimal number of two type-I Weyl points along the Z-$\Gamma$-Z high-symmetry path near the Fermi level without other interfering bands [inset of Fig.~\ref{fig_SOCband}(a)]. The band crossings are mainly caused by the hybridization between the \emph{p} orbitals of the Te and Sb atoms and protected by the \emph{C}$_{3}$ rotational symmetry. Our Wannier charge center (WCC) calculations show that the two Weyl points (WP$_{1}$ and WP$_{2}$) exhibit opposite chirality carrying the topological charges of +1 and -1 [Fig.~\ref{fig_SOCband}(b)], respectively, indicating that MnSn$_{2}$Sb$_{2}$Te$_{6}$-B is an ideal topological WSM. In addition, the calculated Chern numbers $C=1$ at $k_{z}=0$ plane and $C=0$ at $k_{z}=\pi$ plane are also consistent with the feature of an ideal WSM. On the other hand, we also investigated the effect of strain on the topological properties of MnSn$_{2}$Sb$_{2}$Te$_{6}$-B bulk by applying uniaxial stress from -5$\%$ (compression) to 5$\%$ (tensile) along the \emph{c}-axis. As shown in Fig.~\ref{fig_SOCband}(c), for MnSn$_{2}$Sb$_{2}$Te$_{6}$-B bulk in the FM ground state, the energy gap between the valence band maximum (VBM) and the conduction band minimum (CBM) at the $\Gamma$ point increases monotonically with the increase of compressive stress and becomes an FM axion insulator (AXI). However, under tensile stress within 3$\%$ it is still a WSM with a single pair of Weyl points. Once the tensile stress exceeds 3$\%$, the system becomes a trivial FM insulator (FMI). Figure~\ref{fig_SOCband}(d) shows the projected band structure with orbital weights for MnPb$_{2}$Bi$_{2}$Te$_{6}$-B bulk in the FM ground state with an out-of-plane easy magnetization axis, we can see that there is a band inversion between the \emph{p} orbitals of Te and Bi atoms at the $\Gamma$ point near the Fermi level, indicating that the MnPb$_{2}$Bi$_{2}$Te$_{6}$-B bulk may have nontrivial topological properties. For the other five Mn\emph{X}$_{2}$\emph{B}$_{2}$\emph{T}$_{6}$-B bulks in their respective ground states, their band structures calculated with the SOC are shown in Fig. S6 of the SM.

Since the Mn\emph{X}$_{2}$\emph{B}$_{2}$\emph{T}$_{6}$-B family has the inversion symmetry under different ground state magnetic configurations, a parity-based Z$_{4}$ symmetry indicator can be used to characterize the topological properties of Mn\emph{X}$_{2}$\emph{B}$_{2}$\emph{T}$_{6}$-B family. The Z$_{4}$ invariant is given by~\cite{38Turner,39Ono,40Watanabe}
\begin{equation}\label{eq_1}
\begin{aligned}
  Z_{4} &= \Sigma_{\alpha=1}^{8}\Sigma_{n=1}^{n_{occ}}\frac{1+\xi_{n}(\Lambda_{\alpha})}{2}~~mod~4,
\end{aligned}
\end{equation}
where $\Lambda_{\alpha}$ are the eight inversion-symmetric crystal momenta, $\xi_{n}(\Lambda_{\alpha})$ is the parity eigenvalue (+1 or -1) of the \emph{n}th occupied band at the $\Lambda_{\alpha}$, and $n_{occ}$ is the total number of occupied bands. Z$_{4}$=1 or 3 corresponds to a Weyl semimetal (WSM) phase~\cite{38Turner}, while Z$_{4}$=2 indicates an axion insulator (AXI) [the axion angle $\theta=\pi$] in the case of the Chern numbers on all the 2D planes in the BZ are zeros~\cite{41Huan,42Xu}. From Table \ref{tab:II}, we can find that the Z$_{4}$ invariant of MnSn$_{2}$Sb$_{2}$Te$_{6}$-B equals to 1, indicating that it is a WSM, which is consist with our calculated band structures [Fig.~\ref{fig_SOCband}(a)] and WCCs [Fig.~\ref{fig_SOCband}(b)]. While the MnGe$_{2}$Bi$_{2}$Te$_{6}$-B, MnSn$_{2}$Bi$_{2}$Te$_{6}$-B, and MnPb$_{2}$Bi$_{2}$Te$_{6}$-B bulks correspond to Z$_{4}$=2, the calculated Chern numbers at $k_{z}=0$ and $k_{z}=\pi$ planes both equal to 0, implying that they are a class of FM axion insulators. Furthermore, the Z$_{4}$ invariants of MnGe$_{2}$Sb$_{2}$Se$_{6}$-B, MnGe$_{2}$Sb$_{2}$Te$_{6}$-B and MnGe$_{2}$Bi$_{2}$Se$_{6}$-B equal to 0, suggesting that they are a class of trivial insulators.

\begin{figure}[!th]
	\centering
	\includegraphics[width=0.50\textwidth]{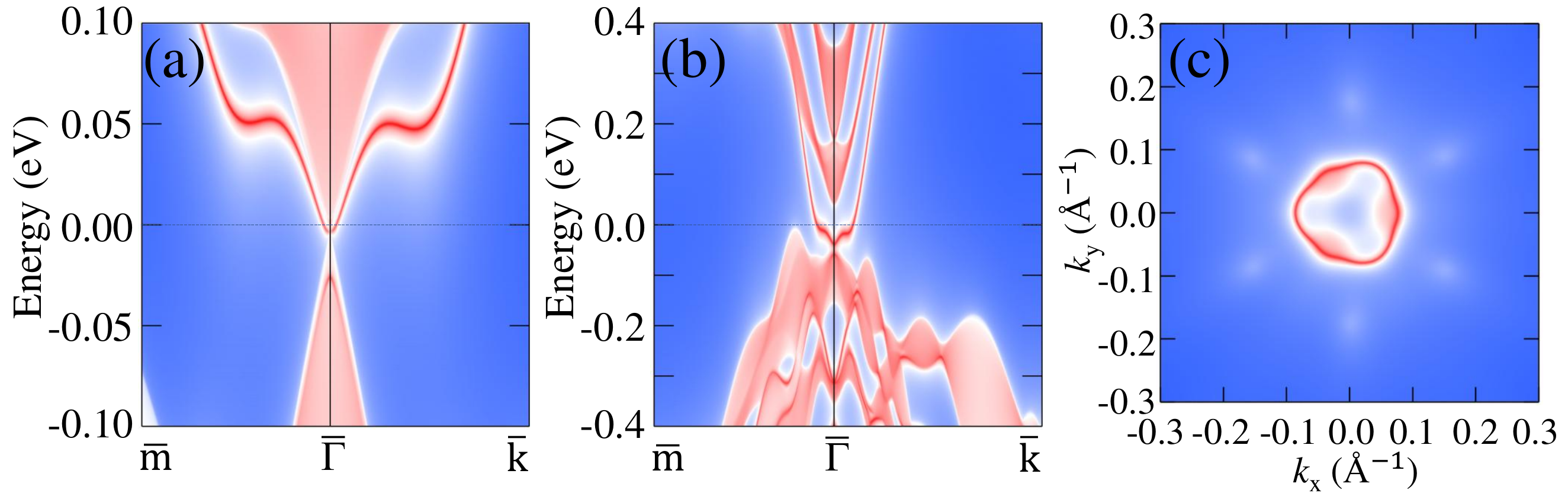}
	\caption{(Color online) Surface energy bands of the semi-infinite (111) surface of (a) MnSn$_{2}$Sb$_{2}$Te$_{6}$-B and (b) MnPb$_{2}$Bi$_{2}$Te$_{6}$-B. (c) A triangular Fermi surface on the (111) surface of MnPb$_{2}$Bi$_{2}$Te$_{6}$.}
	\label{fig_surface}
\end{figure}

Both WSMs and FM axion insulators exhibit unique topological surface states. Considering that Mn\emph{X}$_{2}$\emph{B}$_{2}$\emph{T}$_{6}$-B bulks are formed by ABC stacking of 11 atomic-layers building blocks along the \emph{c}-axis via the vdW interaction, the (111) surface of their primitive cells located in the vdW gap is their natural cleavage plane. So, we calculated the surface states for the (111) surfaces of MnSn$_{2}$Sb$_{2}$Te$_{6}$-B and MnPb$_{2}$Bi$_{2}$Te$_{6}$-B. We can find that the two Weyl points (WP$_{1}$ and WP$_{2}$) of MnSn$_{2}$Sb$_{2}$Te$_{6}$-B bulk are exactly projected to the same point $\bar{\Gamma}$ on the (111) surface, thus the two surface Fermi arcs are displayed at the $\bar{\Gamma}$ point [Fig.~\ref{fig_surface}(a)]. However, for axion insulator, in addition to the above axion angle $\theta=\pi$, a gapped surface state is also required. We take MnPb$_{2}$Bi$_{2}$Te$_{6}$-B as an example to calculate the surface bands in the projected (111) surface [see Fig.~\ref{fig_surface}(b)]. One can see that the surface states at the $\bar{\Gamma}$ point near the Fermi level do open an energy gap ($\sim$20~meV) and are accompanied by a triangular Fermi surface [see Fig.~\ref{fig_surface}(c)], which is similar to the surface state of MnGe$_{2}$Sb$_{2}$Te$_{6}$-A that we previously reported~\cite{18Gao}. Notably, MnPb$_{2}$Bi$_{2}$Te$_{6}$-B has only the gapped surface state at the Fermi level without other interfering bulk states.

$Discussion$: It is well known that the search for the ideal Weyl semimetals with a single pair of Weyl points and the intrinsic FM axion insulators in magnetic materials with highly desirable topological states has been a challenging task. Despite a long search, their candidates are still rare so far~\cite{17Hu,18Gao,19Nie}. Fortunately, both of the two unique topological states can be found in the Mn\emph{X}$_{2}$\emph{B}$_{2}$\emph{T}$_{6}$-B family. Compared with our previous work on Mn\emph{X}$_{2}$\emph{B}$_{2}$\emph{T}$_{6}$-A, the Mn\emph{X}$_{2}$\emph{B}$_{2}$\emph{T}$_{6}$-B family has the following advantages. First, not only the energy of the Mn\emph{X}$_{2}$\emph{B}$_{2}$\emph{T}$_{6}$-B family is lower than that of the Mn\emph{X}$_{2}$\emph{B}$_{2}$\emph{T}$_{6}$-A family, but also the former is the stable structural form of Mn\emph{X}$_{2}$\emph{B}$_{2}$\emph{T}$_{6}$ system. Second, a new topological phase emerges from the Mn\emph{X}$_{2}$\emph{B}$_{2}$\emph{T}$_{6}$-B family, that is, the ideal magnetic Weyl state with only a single pair of Weyl points. Third, MnPb$_{2}$Bi$_{2}$Te$_{6}$-B has only the gapped surface state at the Fermi level without other interfering bulk states. Thus, it can provide a more ideal platform for future experiments to explore the long-sought quantized magnetoelectric effect based on the FM axion insulators. Finally, we emphasize that our design scheme can provide new ideas for realizing more vdW-type magnetic topological materials.

$Summary$: To summarize, we propose a new class of Mn\emph{X}$_{2}$\emph{B}$_{2}$\emph{T}$_{6}$-B (\emph{X}=Ge, Sn, or Pb; \emph{B}=Sb or Bi; \emph{T}=Se or Te) family that is the stable structural form of the Mn\emph{X}$_{2}$\emph{B}$_{2}$\emph{T}$_{6}$-A materials we reported before. We systematically investigate the stability, magnetic, electronic and topological properties of the Mn\emph{X}$_{2}$\emph{B}$_{2}$\emph{T}$_{6}$-B family and find that Mn\emph{X}$_{2}$\emph{B}$_{2}$\emph{T}$_{6}$-B monolayers are narrow-bandgap ferromagnetic (FM) semiconductors in their FM ground state with an out-of-plane easy magnetization axis, while Mn\emph{X}$_{2}$\emph{B}$_{2}$\emph{T}$_{6}$-B bulks not only have the intrinsic FM axion insulators MnGe$_{2}$Bi$_{2}$Te$_{6}$-B, MnSn$_{2}$Bi$_{2}$Te$_{6}$-B, and MnPb$_{2}$Bi$_{2}$Te$_{6}$-B, but also the intrinsic WSM MnSn$_{2}$Sb$_{2}$Te$_{6}$-B with only a single pair of Weyl points that has not appeared in the Mn\emph{X}$_{2}$\emph{B}$_{2}$\emph{T}$_{6}$-A family. Thus, the new Mn\emph{X}$_{2}$\emph{B}$_{2}$\emph{T}$_{6}$-B family can provide an ideal platform for future experiments to explore the long-sought quantized magnetoelectric effect and the intrinsic properties related to Weyl points.

\begin{acknowledgments}
We wish to thank QuanSheng Wu and Wei Liu for helpful discussions. This work was supported by the National Key R\&D Program of China (Grant No. 2019YFA0308603), the National Natural Science Foundation of China (Grants no. 11934020 and No. 12174443), and the Beijing Natural Science Foundation (Grant No. Z200005). Y. G is also grateful for the support of HZWTECH for providing computational facilities.
\end{acknowledgments}

\end{document}